\documentclass[pre,twocolumn,groupedaddress, showkeys,showpacs]{revtex4}
\usepackage{graphicx}
\begin{document}

%\preprint{}

\title{Double negative differential thermal resistance induced by the nonlinear on-site potentials}


\author{Bao-quan  Ai$^{1}$}%\email[Email: ]{aibq@hotmail.com}
\author{Wei-rong Zhong$^{2}$} \email[Email: ]{wrzhong@jnu.edu.cn}
\author{Bambi Hu$^{3,4}$}

%\homepage[]{}

%\thanks{}
\affiliation{$^{1}$ Laboratory of Quantum Information Technology, ICMP and
 SPTE, South China Normal University, Guangzhou, China.\\
 $^{2}$Department of Physics, College of Science and Engineering,
Jinan University, 510632 Guangzhou, China.\\
$^{3}$ Department of Physics, Centre for Nonlinear Studies, and the
Beijing-Hong Kong-Singapore Joint Centre for
Nonlinear and Complex Systems (Hong Kong), Hong Kong Baptist University, Kowloon Tong, Hong Kong, China\\
 $^{4}$Department of Physics, University of Houston, Houston, Texas 77204-5005, USA}


%\date{\today}
\begin{abstract}
We study heat conduction through one-dimensional homogeneous
lattices in the presence of the nonlinear on-site potentials
containing the bounded and unbounded parts, and the harmonic
interaction potential. We observe the occurrence of double negative
differential thermal resistance (NDTR), namely, there exist two
regions of temperature difference, where the heat flux decreases as
the applied temperature difference increases. The nonlinearity of
the bounded part contributes to NDTR at low temperatures and NDTR at
high temperatures is induced by the nonlinearity of the unbounded
part.  The nonlinearity of the on-site potentials is necessary to
obtain NDTR for the harmonic interaction homogeneous lattices.
However, for the anharmonic homogeneous lattices, NDTR even occurs
in the absence of the on-site potentials, for example the rotator
model.
\end{abstract}

\pacs{05.70.Ln, 44.10.+i, 05.60.-k}
\keywords{Double negative differential thermal resistance, heat conduction }



\maketitle
%\section {Introduction}
\indent Heat conduction in low-dimensional systems has become the
subject of a large number of theoretical and experimental studies in
recent years \cite{a1}. The theoretical interest in this field lies
in the rapid progress in probing and manipulating thermal properties
of nanoscale systems, which unveils the possibility of designing
thermal devices with optimized performance at the atomic scale. As
we all know, devices that control the transport of electrons, such
as the electrical diode and transistor, have been extensively
studied and led to the widespread applications in modern
electronics. However, it is far less studied for their thermal
counterparts as to control the transport of phonons (heat flux),
possibly by reason that controlling phonons is more difficult than
controlling electrons. Recently, it has been revealed by theoretical
studies in model systems that,
 such as electrons and photons, phonons can also perform interesting function
\cite{a2}, which shed light on the possible designs of thermal
devices. For example, heat conduction in asymmetric nonlinear
lattices demonstrates rectification phenomenon, namely, the heat
flux can flow preferably in a certain direction
\cite{a3,a4,a5,a6,a7,a8,a9,a10}.  Remarkably, a thermal rectifier
has been experimentally realized by using gradual mass-loaded carbon
and boron nitride nanotubes \cite{a11}. The nonlinear systems with
structural asymmetry can exhibit thermal rectification, which has
triggered model designs of various types of thermal devices such as
thermal transistors \cite{a8}, thermal logic gates \cite{a12}, and
thermal memory\cite{a13}. It is worth pointing out that most of
these studies are relevant to heat conduction in the nonlinear
response regime, where the counterintuitive phenomenon of NDTR may
be observed and plays an important role in the operation of those
devices.

\indent NDTR refers to the phenomenon where the resulting heat flux
decreases as the applied temperature difference (or gradient)
increases. It can be seen that a comprehensive understanding of the
phenomenon of NDTR, which is lacking at the moment, would be
conducive to further developments in the designing and fabrication
of thermal devices.  The existing studies on NDTR have been on
models with structural inhomogeneity, such as the two-segment
Frenkel-Kontorova model \cite{a8,a14}, the weakly coupled
two-segment $\phi^{4}$ model \cite{a15}, and the anharmonic graded
mass model \cite{a6}. However, structural asymmetry is not a
necessary condition for NDTR.  In the nonlinear response regime,
NDTR can occur in absolutely symmetric structures where there exists
nonlinearity in the on-site potential of the lattice model
\cite{a16}.  However, it is still not clear whether multiple NDTR
can occur in symmetric structures. In this Brief Report, we study
the exhibition of double NDTR in the absolutely symmetric structures
and find the occurrence of double NDTR. Furthermore, we also find
that NDTR can also occur in the coupled rotator model where the
on-site potential is absent.

%\section {Model and methods}
\indent In this study, the homogeneous lattice models are each
described by a Hamiltonian of the form
\begin{equation}\label{}
    H=\sum
    _{i=1}^{N}\frac{p_{i}^{2}}{2m}+V(x_{i}-x_{i+1})+U(x_{i}),
\end{equation}
where $p_{i}$ is the momentum of the $i$th particle, and $x_{i}$ its
displacement from equilibrium position. $m$ is the mass of the
particles. $V(x)$ is the nearest-neighbor interaction potential, the
harmonic potential is used,
\begin{equation}\label{}
    V(x)=\frac{1}{2}kx^{2},
\end{equation}
where $k$ is the coupling constant. As for the on-site potential
$U(x)$, we consider two cases shown in Fig. 1. For case $A$ ($\phi^{4}$ model)\cite{a1}
\begin{equation}\label{}
    U(x)=-\frac{\alpha}{2}x^{2}+\frac{\lambda}{4}x^{4},
\end{equation}
and for case $B$
\begin{equation}\label{}
    U(x)=-\frac{U_{0}}{(2\pi)^{2}}\cos(2\pi
    x)+\frac{\lambda}{4}x^{4},
\end{equation}
where $\alpha$, $U_{0}$, and $\lambda$ are the parameters that
control the shape of the potential. The on-site potential contains
the bounded and unbounded parts.

\begin{figure}[htbp]
\begin{center}\includegraphics[width=7cm,height=8cm]{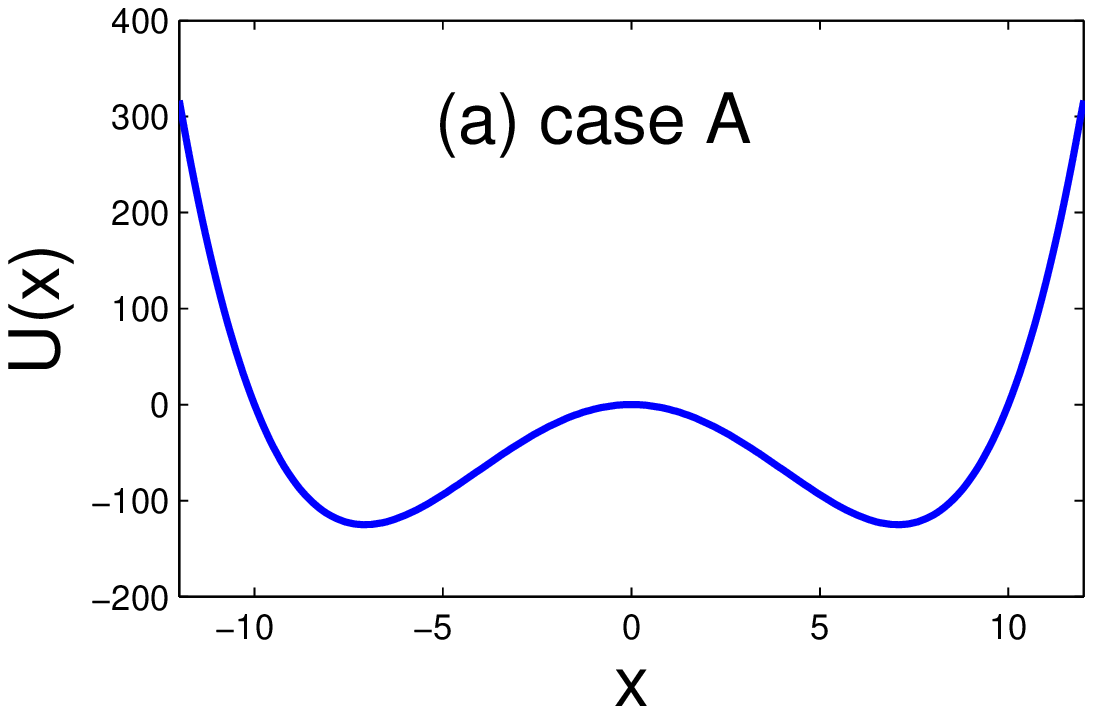}
\includegraphics[width=8cm,height=7cm]{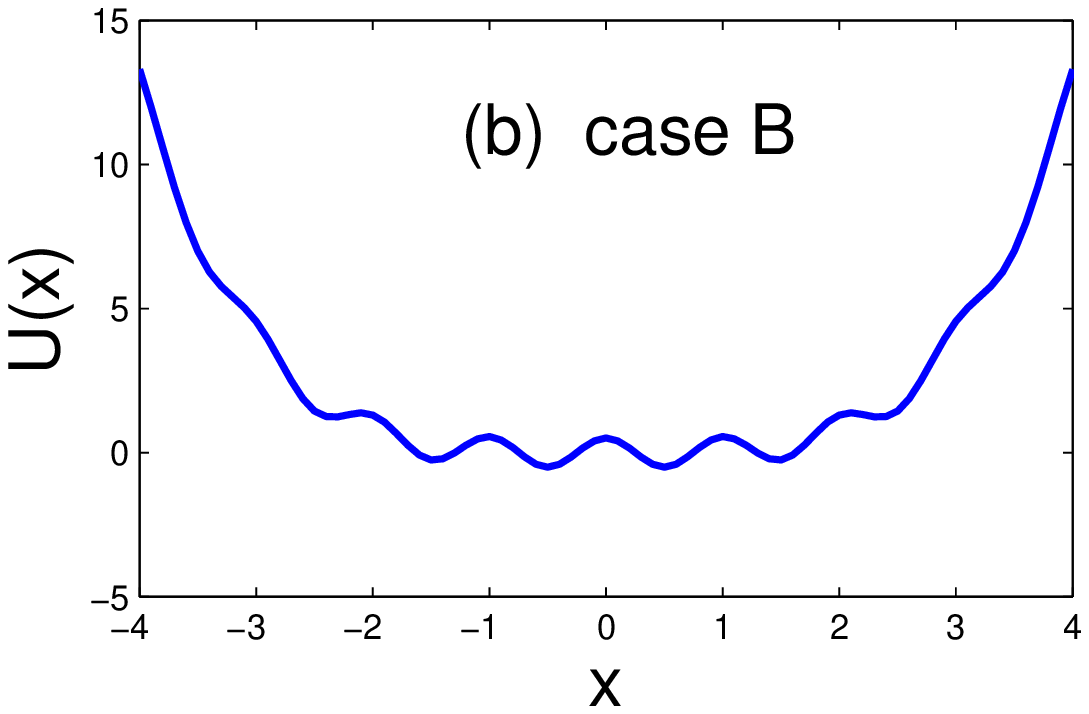}
\caption{(Color online) The shape of on-site potential. (a) Case $A$:
$U(x)=-\frac{\alpha}{2}x^{2}+\frac{\lambda}{4}x^{4}$ ($\alpha>0$),
(b) case $B$: $U(x)=\frac{U_{0}}{(2\pi)^{2}}\cos(2\pi
    x)+\frac{\lambda}{4}x^{4}$. }\label{1}
\end{center}
\end{figure}

As to obtain a stationary heat flux, the chain is connected two heat
baths at temperature $T_{+}$ and $T_{-}$, respectively. Fixed
boundary conditions are taken $x_{0}=x_{N+1}=0$. For each of the
one-dimensional lattice models, the equation of motion takes the
form
\begin{equation}\label{}
    m\ddot{x}_{i}=-\frac{\partial H}{\partial x_{i}}-\gamma_{i}\dot{x}_{i}+\xi_{i},
\end{equation}
where $\gamma_{i}=\gamma(\delta_{i,1}+\delta_{i,N})$ and
$\xi_{i}=\xi_{+}\delta_{i,1}+\xi_{-}\delta_{i,N}$. The noise terms
$\xi_{\pm}(t)$ denote a Gaussian white noise that has a zero mean
and a variance of $2\gamma k_{B}T_{\pm}$, where $\gamma$ is the the
friction coefficient and $k_{B}$ is Boltzmann's constant. The dot
stands for the derivative operator with respect to time $t$. The
local heat flux is given by $j_{i}=\langle
\dot{x_{i}}F(x_{i}-x_{i-1})\rangle$, where $F=-\frac{\partial
V}{\partial x}$ and the notation $\langle. . . \rangle$ denotes a
steady-state average.  The local temperature is defined as
$T_{i}=\langle m\dot{x_{i}}^{2}\rangle$. After the system reaches a
stationary state, $j_{i}$ is independent of site position $i$, so
that the flux can be denoted as $j$. In our simulations, the
equations (5)
 of motion are integrated by using a second order stochastic Runge-Kutta algorithm\cite{b1} with a small
time step ($h=0.001$).  The simulations are performed long enough to
allow the system to reach a nonequilibrium steady state in which the
local heat flux is a constant along the chain.

%\section {Numerical results and discussion}
\begin{figure}[htbp]
\begin{center}\includegraphics[width=7cm]{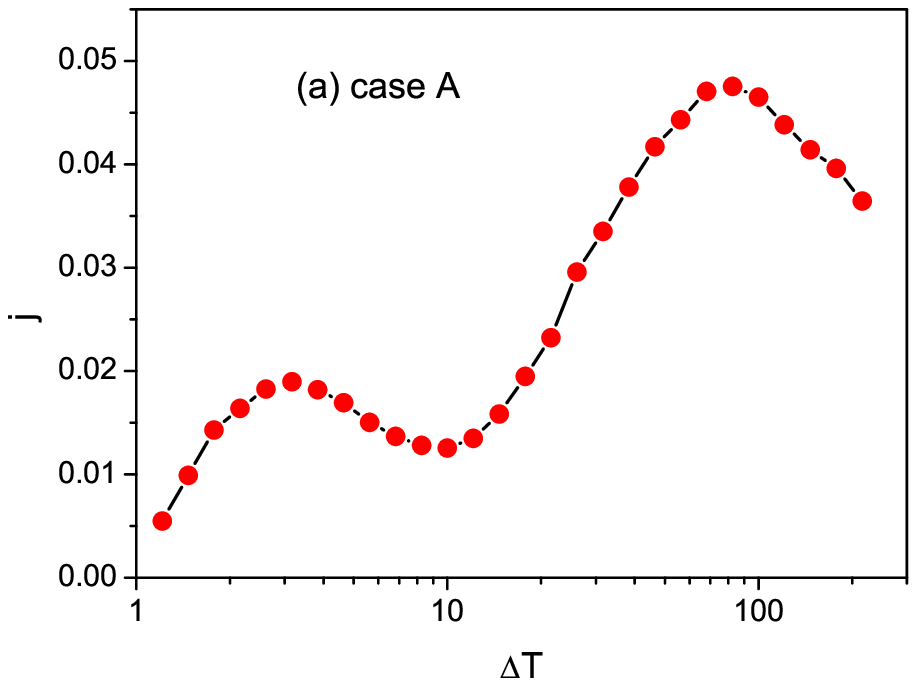}
\includegraphics[width=7cm]{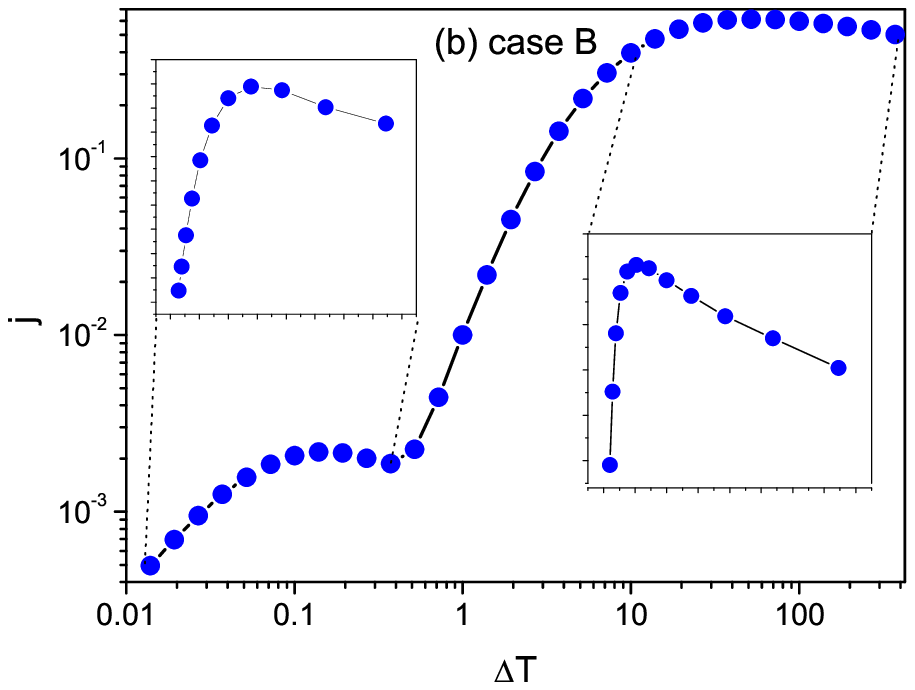}
\caption{(Color online) Heat flux $j$ as a function of temperature difference
$\Delta T$ for case $A$ and $B$. (a) For case $A$: $\alpha=10$,
$\lambda=2.0$, and $T_{-}=0.01$; (b) for case $B$: $U_{0}=10.0$,
$\lambda=0.2$, $T_{-}=0.001$, and $N=32$. The insets in (b) give the
enlarged views of the NDTR.  The other parameters are $k=1.0$,
$N=32$, $T_{+}=T_{-}+\Delta T$.}\label{1}
\end{center}
\end{figure}

\indent Figure 2 (a) shows the dependence of the heat flux $j$ on
temperature difference $\Delta T$ for $\phi^{4}$ model with
$\alpha>0$ (case A). It is found that there exist two regions of
$\Delta T$, in which the larger temperature difference the less heat
flux through the system, namely double NDTR occurs. The presence of
a nonlinear on-site potential facilitates the occurrence of
phonon-lattice scattering, which generally becomes more significant
for increasing temperature and can therefore contribute to a
decrease in the thermal conductivity. For low temperatures (small
temperature difference), the bounded part of the on-site potential
dominates the system. For this case, the phonon-lattice scattering
is important only at sufficiently low temperatures where the
dynamics of the particles is much influenced by the bounded on-site
potential. As the applied temperature difference $\Delta T$
increases from zero with $T_{-}$ being fixed, the increase in the
thermodynamic driving force will drive an increase in the heat flux.
At higher values of $\Delta T$, however, the effect of
phonon-lattice scattering becomes so significant that the first NDTR
occurs. But with a further increase in $\Delta T$, the average
temperature of the system has become sufficiently high such that the
particles can overcome the bounded part of on-site potential, the
phonon-lattice scattering becomes not significant and the first NDTR
disappears. At the same time, the dynamics of the particles is
dominated by the unbounded part of the on-site potential.  As
increasing the temperature, the increase in phonon-lattice
scattering is reflected by the power-law decrease of the thermal
conductivity and the second NDTR occurs. Therefore, we can obverse
the double NDTR as the temperature difference increases from zero
with $T_{-}$ being fixed. Form Fig. 2 (b), we can see that the
similar double NDTR can also occur for case B.

\indent Since the results for case A are similar to that for case
B, we will only focus on case B for investigating the parameters
dependence of double NDTR.  The on-site potential of case B is
composed of two parts: the bounded part
$-\frac{U_{0}}{(2\pi)^{2}}\cos(2\pi x)$ and unbounded part $\frac{\lambda}{4}x^{4}$.

\begin{figure}[htbp]
\begin{center}\includegraphics[width=7cm]{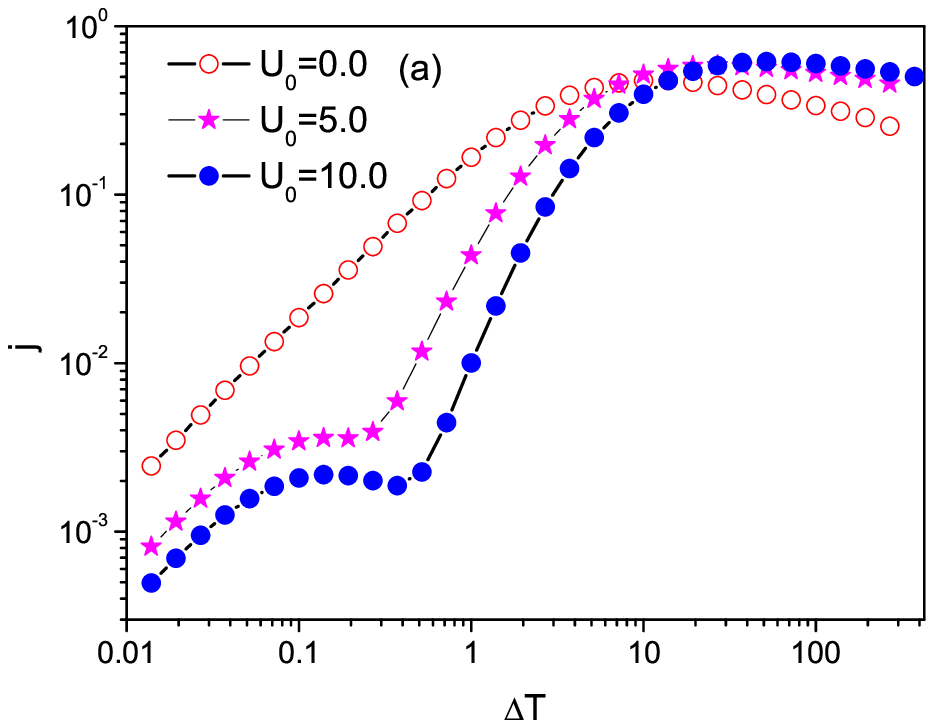}
\includegraphics[width=7cm]{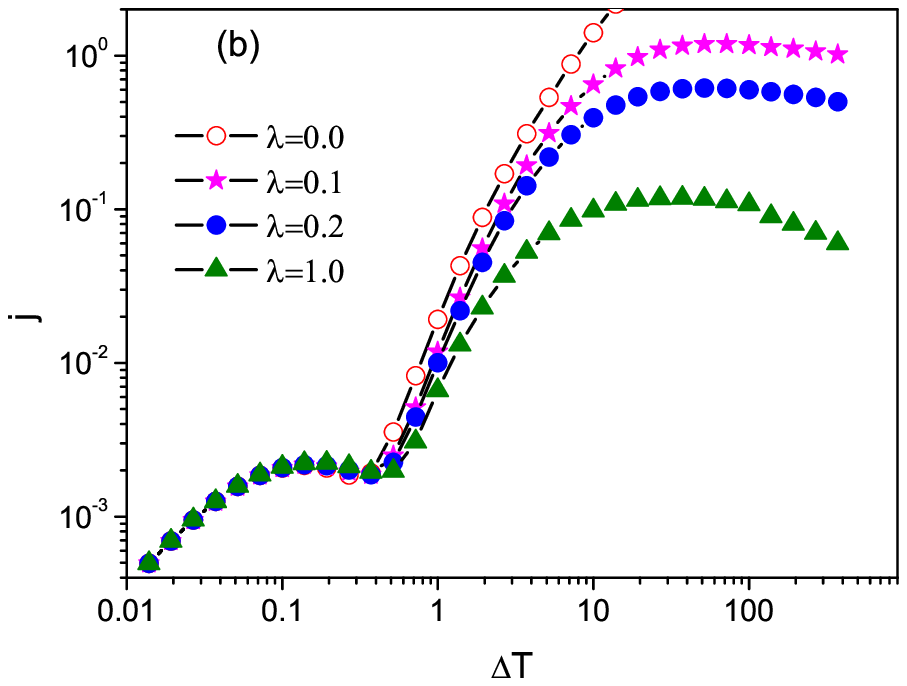}
\includegraphics[width=7cm]{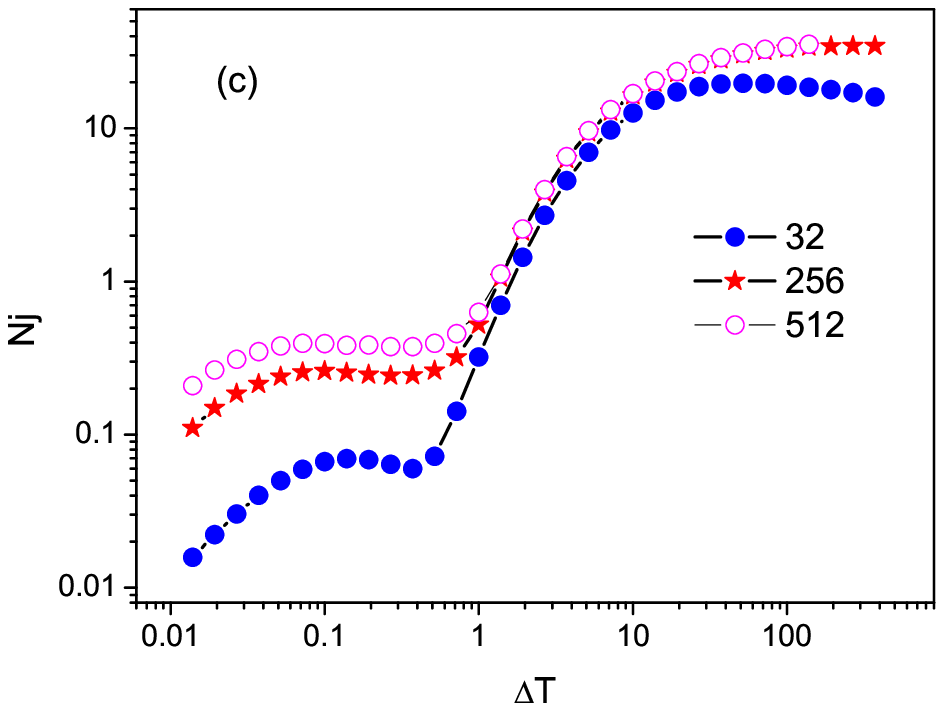}
\includegraphics[width=7cm]{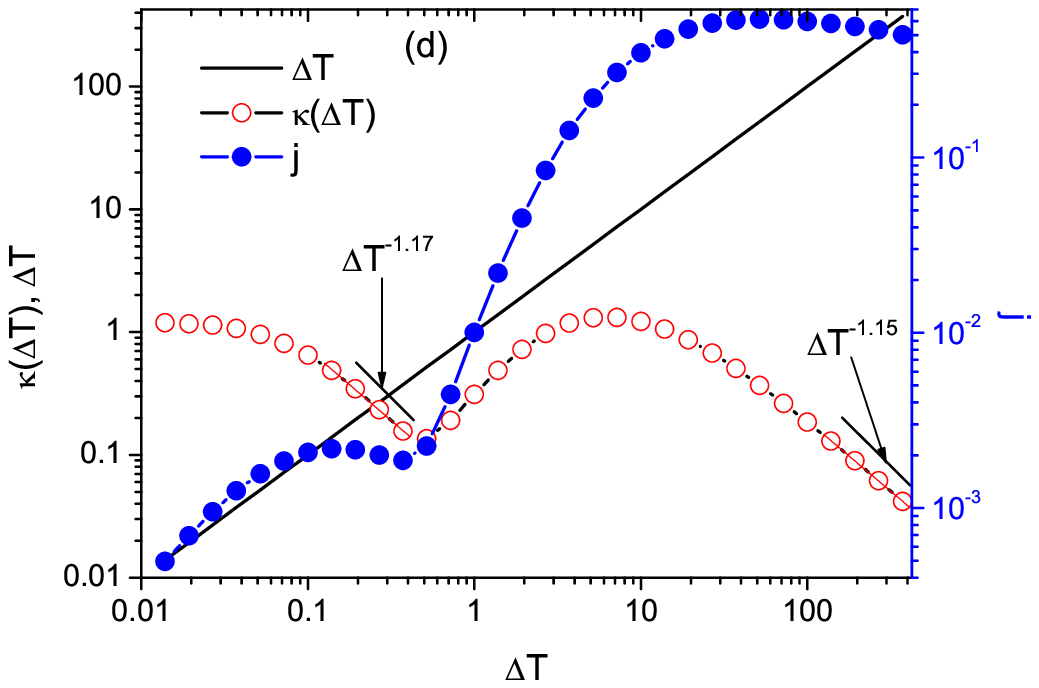}
\caption{(Color online) (a) Heat flux $j$ as a function of $\Delta
T$ for different values of $U_{0}$ at $k=1.0$, $\lambda=0.2$, and
$N=32$; (b) Heat flux $j$ as a function of $\Delta T$ for different
values of $\lambda$ at $k=1.0$, $U_{0}=10.0$, and $N=32$; (c) The
total heat flux $Nj$ as a function of $\Delta T$ for $N=32$, $256$,
and $512$ at $U_{0}=10.0$, $k=1.0$, and $\lambda=0.2$; (d) $\kappa
(\Delta T)$, $\Delta T$, and heat flux $j$ as a function of $\Delta
T$ at $U_{0}=10.0$, $k=1.0$, $\lambda=0.2$, and $N=32$. The other
parameters are $T_{-}=0.001$, $T_{+}=T_{-}+\Delta T$. }\label{1}
\end{center}
\end{figure}
\indent Figure 3 (a) shows the heat flux $j$ as a function of
$\Delta T$ for different values of $U_{0}$ with $\lambda$ being
fixed. For decreasing $U_{0}$, the first NDTR region
becomes smaller, and finally disappears.  Therefore, the bounded
part of the on-site potential contributes to the occurrence of the
first NDTR.  From Fig. 3 (b), one can find that the occurence of the
second NDTR is induced by the unbounded part of the on-site
potential. Figure 3 (c) shows that the two NDTR regimes generally
become smaller as the system size $N$ increases, and eventually
vanish in the thermodynamic limit.

\indent Figure 3 (d) shows the effective thermal conductivity
$\kappa(\Delta T)$, heat flux $j$, and $\Delta T $ as a function of
$\Delta T$ at $U_{0}=10.0$, $k=1.0$, $\lambda=0.2$, and $N=32$.  As
we know $j=\kappa(\Delta T)\Delta T$. It is found that
$\kappa(\Delta T)\propto \Delta T^{-1.17}$ for the first NDTR
region, and $\kappa(\Delta T)\propto \Delta T^{-1.15}$ for the
second NDTR region, resulting in $j\propto T^{-0.17}$ for the first
NDTR region, and $j\propto T^{-0.15}$ for the second NDTR region.

\begin{figure}[htbp]
\begin{center}\includegraphics[width=7cm]{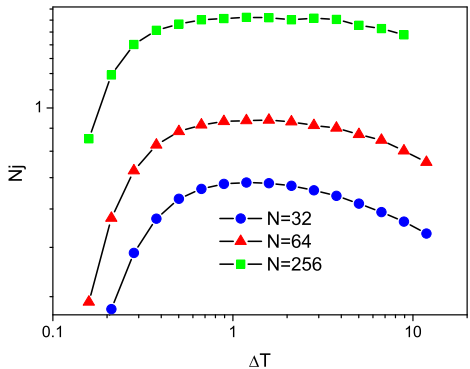}
\caption{(Color online) The total heat flux $Nj$ as a function of
temperature difference $\Delta T$.  The other parameters are
$T_{-}=0.1$, $T_{+}=T_{-}+\Delta T$.}\label{1}
\end{center}
\end{figure}

\indent  Note that there is no definite relation between the shape
of the on-site potentials and the occurrence of NDTR. For example,
if the term $x^{4}$ in case B is replaced by the harmonic term
(linear potential) $x^{2}$, the shape of the on-site potentials does
not change essentially,  while NDTR at high temperatures will
disappear. Therefore, it is the nonlinearity, not the shape of the
on-site potentials,  that determines the occurrence of NDTR.  The
nonlinearity in the on-site potentials is necessary to obtain NDTR
for the harmonic interaction homogeneous systems.  Now we return to
the anharmonic homogenous systems and check if NDTR can occur in the
absence of the on-site potentials. From the previous work
\cite{a16}, one can find that NDTR can not occur in pure harmonic
and Fermi-Pasta-Ulam model. However, in that work, the rotator model
was not considered. The simplest example of rotator model with
nearest neighbor interactions lies in the class (1):
$V(x)=1-\cos(x)$ and $U(x)=0$. From Fig. 4, we can see that NDTR
occurs in the rotator model, which is related to excitation of
nonlinear localized rotational modes of the chain\cite{b2}. In
addition, NDTR can also occur in the anharmonic graded mass lattices
\cite{a6}.  Therefore, for the anharmonic systems, the on-site
potential is not a necessary condition for NDTR.

%\section{Concluding Remarks}
\indent In conclusion, we study heat conduction through the
one-dimensional homogeneous lattices with the nonlinear on-site
potentials. The on-site potentials are composed of two parts: the
bounded and unbounded parts. From nonequilibrium molecular dynamics
simulations, it is found that double NDTR occurs as the temperature
difference increases.  The occurrence of NDTR at low temperatures is
caused by  the nonlinearity of the bounded part, while the
nonlinearity of the unbounded part induces the occurrence of NDTR at
high temperatures. In addition, we also find that NDTR even occurs
in anharmonic homogenous lattices without on-site potentials, for
example the rotator model. Therefore, the on-site potential is not
necessary to obtain NDTR for the anharmonic lattices. It is also
found that the regime of NDTR becomes smaller as the system size
increases, and eventually vanishes in the thermodynamic limit. The
observation of double NDTR in homogeneous systems shows that the
nonlinearity of the on-site potentials is very important for
designing NDTR devices. It is possible to design the thermal devices
with the more complex functions by using the occurrence of double
NDTR.

  \indent This work was supported in part by National
Natural Science Foundation of China (Grant Nos. 30600122, 11004082
and 10947166 )and GuangDong Provincial Natural Science Foundation
(Grant No. 06025073 and 01005249).

\end{document}